\documentclass[amssymb,aps,notitlepage,showpacs,showkeys]{revtex4-1}
\usepackage{graphicx}
\usepackage{fancyvrb}

\usepackage{color}
\definecolor{brown}{rgb}{0.8,0.6,0.3}
\definecolor{dgreen}{rgb}{0.2,0.4,0.3}

\usepackage[pdfauthor={Richard J. Mathar}, pdftitle={arXiv:2004.11808}, colorlinks=true,urlcolor=blue,linkcolor=red,citecolor=brown,pdfkeywords={{Ground-based Astronomy},{Atmospheric Dispersion},{Scale Height}},pdftitle={Atmospheric Dispersion: Exponential Model},bookmarksopen,pdfpagelayout=OneColumn]{hyperref}

\begin{document}

\title[Barometric Model of Refractive Index]{
A Barometric Exponential Model of the Atmosphere's Refractive Index: Zenith Angles and Second Order Aberration in the Entrance Pupil
}

\author{Richard J. Mathar}
\homepage{https://www.mpia-hd.mpg.de/homes/mathar}
\affiliation{
Max-Planck Institute of Astronomy, K\"onigstuhl 17, 69117 Heidelberg, Germany}
\pacs{97.75.Pq, 42.68.Wt}
\keywords{Atmospheric Refraction, Telescope Focus, Optical Path Difference}

\date{\today}

\begin{abstract}
This report models the refractive index above a telescope site
by an atmosphere with exponential decay
of the refractive index (susceptibility) as a function of altitude. The air is represented
as a spherical hull around the surface.
We compute (i) the differential zenith angle --- the difference between
the actual zenith angle at arrival of rays at the telescope and a
hypothetical zenith angle without the atmosphere --- and (ii) the optical path
length distribution of rays at arrival in the entrance pupil as a function of
the distance to the pupil center.

The key technique in this work is to expand
some integrals --- that depend on the refractive index profile along the
curved path of each ray from some virtual plane in the direction
of the star up to the entrance pupil --- in power series of small parameters.
\end{abstract}

\maketitle

\section{Spherical Symmetry of the Atmospheric Hull}
\begin{figure}
\includegraphics[scale=0.4]{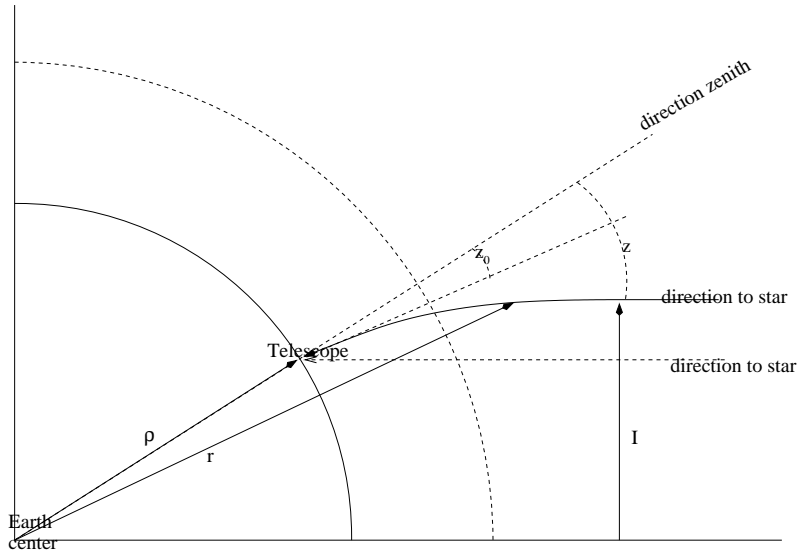}
\caption{Distance $\rho$ of the telescope to the Earth center, distance $r$ of a point on
the curved path of the ray, true zenith angle $z_0$, astrometric zenith angle $z$, 
and impact parameter $I$. (Not to scale.)}
\label{fig.sph}
\end{figure}

In ground-based astronomy, light emitted
by astronomical objects is tilted more toward the zenith axis
when arriving at the telescope than computed by standard spherical astronomy;
the atmosphere acts like a gradient lens, and the actual zenith angle of arrival
$z_0$ differs from the astrometric zenith angle $z$ of a hypothetical Earth in vacuum
by an angle 
\begin{equation}
R=z-z_0. 
\label{eq.R}
\end{equation}
Figure \ref{fig.sph} sketches the relevant variables.
Applying successively Snell's equations
as light changes its direction when it encounters gradients
in the refractive index $n$, the angle $R$ is a functional
of the refractive index along the path 
\cite{MahanAO1,Green1985,ThomasJHUAPL17,AuerAJ119,NenerJOSA20,NoerdlingerJPRS54,Tannousarxiv01}
\begin{equation}
R=\rho n_0 \sin z_0\int_1 ^{n_0} \frac{dn}{n\sqrt{r^2n^2-\rho^2 n_0^2\sin^2 z_0}}
\end{equation}
measured in radians.
$\rho\approx 6378$ km is the radius of the Earth representing the curvature of the
atmosphere above the telescope \cite{NIMA8350}. 
$n(r)$ is the refractive index at a distance $r\ge \rho$ to
the Earth center (!); it equals unity high above the atmosphere
and increases to $n_0>1$ at the telescope's altitude.

Evaluating $R$ is a key to faithful pointing models and all-sky mapping
of astrometry. Inserting wavelength-dependent refractive indices $n$
leads to designs of atmospheric dispersion correctors.

We start from the Taylor expansion of $R$ in odd powers of the tangent
of the observed zenith angle \cite{StonePASP108,StonePASP114},
\begin{equation}
R\equiv \sum_{i=1,3,5,\ldots} \gamma_i\tan^i z_0,
\label{eq.gammidef}
\end{equation}
such that \cite{MatharBA14}
\begin{eqnarray}
\gamma_1 &=& \rho n_0 \int_1^{n_0} \frac{1}{n^2r}dn;\\
\gamma_3 &=& \frac12 \rho n_0 \int_1^{n_0} \frac{\rho^2n_0^2-r^2n^2}{n^4r^3}dn;\\
\gamma_5 &=& \frac38 \rho n_0 \int_1^{n_0} \frac{(\rho^2n_0^2-r^2n^2)^2}{n^6r^5}dn;\\
\gamma_i &=& {{i-1}\choose{\lfloor i/2\rfloor}}\frac{1}{4^{\lfloor i/2 \rfloor}} 
              \rho n_0 \int_1^{n_0} \frac{(\rho^2n_0^2-r^2n^2)^{\lfloor i/2\rfloor}}{n^{i+1}r^i}dn,\quad i =1,3,5,\ldots\, \mathrm{odd},
\end{eqnarray}
where $\lfloor . \rfloor$ is the floor function, the largest integer not larger than the term inside the bracket.

\section{Exponential Refractivity of Altitude}\label{sec.expo}
The first theme of this manuscript is to derive manageable formulas
for the coefficients $\gamma_i$ by introducing an exponential model
of the refractive index $n(r)$ as a function of altitude \cite{BerberanAJP65,WestJAP81}.
The refractive index is related to the susceptibility $\chi$ via
\begin{equation}
n(r)=\sqrt{1+\chi(r)}
\end{equation}
and we assume that the susceptibility is essentially proportional to an air density
of scale height $K$ such that
\begin{equation}
\chi(r) = \chi_0 e^{-(r-\rho)/K}.
\end{equation}
[Note that this is not Fletcher's model \cite{FletscherMNRAS91} which has another
term proportional to $e^{-(r-\rho)/K}/(r-\rho)$.]
At typical altitudes of observatories the susceptibility at the telescope
is $\chi_0\approx 4\times 10^{-4}$, and the Earth atmosphere has
a scale height $K\approx 9.6$ km.

With such an explicit refractive
index profile, refractive indices are mapped onto dimensionless altitudes 
$\zeta\equiv (r-\rho)/K$:
\begin{equation}
\frac{dn}{d\zeta}=\frac{1}{2n}\frac{d\chi}{d\zeta},
\end{equation}
\begin{equation}
\frac{d\chi}{d\zeta} = -\chi(\zeta),
\end{equation}
and so the variables in the integrals of the previous chapter turn into:
\begin{eqnarray}
\gamma_1 &=& \rho n_0 \int_0 ^\infty \frac{1}{n^2(\rho+\zeta K)}\frac{1}{2n}\chi(\zeta) d\zeta;\\
\gamma_3 &=& \rho n_0 \frac12  \int_0^\infty \frac{\rho^2n_0^2-r^2n^2}{n^4(\rho+\zeta K)^3}\frac{1}{2n}\chi(\zeta)d\zeta;\\
\gamma_5 &=& \rho n_0 \frac38  \int_0^\infty \frac{(\rho^2n_0^2-r^2n^2)^2}{n^6(\rho+\zeta K)^5}\frac{1}{2n}\chi(\zeta)d\zeta.
\end{eqnarray}
The associated
bivariate power series of $\gamma_1$ in terms of the small $\chi_0$ and $K/\rho$ is
written down up to mixed third order in these two variables and integrated term-by-term:
\begin{eqnarray}
\gamma_1 
&=& \rho n_0 \frac{1}{2} \int_0^\infty \frac{1}{n^3(\rho+\zeta K)}\chi_0 e^{-\zeta} d\zeta \nonumber\\
&=& n_0 \chi_0 \frac{1}{2} \int_0^\infty \frac{1}{(1+\chi_0 e^{-\zeta})^{3/2}(1+\zeta K/\rho)} e^{-\zeta} d\zeta 
\nonumber\\
&=& n_0 \chi_0 \frac{1}{2} \int_0^\infty e^{-\zeta} \big[1
-\frac32 e^{-\zeta}\chi_0
-\zeta \frac{K}{\rho}
+\frac{15}{8} e^{-2\zeta}\chi_0^2
+\frac{3}{2} \zeta e^{-\zeta}\chi_0 \frac{K}{\rho}
+\zeta^2 (\frac{K}{\rho})^2
 \nonumber\\
&&
+\frac{35}{16}e^{-3\zeta}\chi_0^3
-\frac{15}{8}\zeta e^{-2\zeta}\chi_0^2 \frac{K}{\rho}
-\frac{3}{2}\zeta^2 e^{-\zeta}\chi_0 (\frac{K}{\rho})^2
-\zeta^3 (\frac{K}{\rho})^3\pm \cdots
\big]d\zeta \nonumber\\
&=& n_0 \chi_0 \frac{1}{2} \left[1-\frac{3}{4}\chi_0 -\frac{K}{\rho}+\frac58 \chi_0^2+\frac38 \chi_0\frac{K}{\rho}
+2(\frac{K}{\rho})^2
-\frac{35}{64}\chi_0^3
-\frac{5}{24}\chi_0^2\frac{K}{\rho}
-\frac{3}{8}\chi_0(\frac{K}{\rho})^2
-6(\frac{K}{\rho})^3
\pm \cdots\right]
.
\label{eq.chi1}
\end{eqnarray}
This kind of expansion as a power series of $K/\rho$ is not new \cite{EshahgAS43}.
\begin{itemize}
\item
In practise $\chi_0$ is known not better than a relative accuracy of $10^{-4}$,
and $K/\rho$ is of the same magnitude;
so keeping more than the first 3 terms in the square bracket in that expansion is a mere academic exercise
from the point of view of an error budget.
\item
If gas mixtures of the air are treated as independent linear responses such that $\chi_0$ is 
an additive composition proportional to the weighted contribution of the air components, this formula does obviously
not apply beyond the first terms in the square bracket, because otherwise mixed quadratic
terms of the individual $\chi_0$ of the constituents interfere with the higher orders (App.\ \ref{app.segr}). 
\item
At a precision of better than a milli-arcsecond,
the oblate ellipsoidal Earth model introduces two different
principal curvatures $1/\rho$ here that
induce an azimuthal wobble of the terms 
\cite{MatharArxiv0907}.
\end{itemize}

For the third order in $\tan z_0$ up to second mixed order
\begin{eqnarray}
\gamma_3 
&=& \rho n_0 \frac{1}{4} \int_0^\infty \frac{\rho^2n_0^2-(\rho+\zeta K)^2n^2}{n^5(\rho+\zeta K)^3}\chi_0 e^{-\zeta} d\zeta
\nonumber\\
&=& n_0 \chi_0 \frac{1}{4} \int_0^\infty \frac{n_0^2-(1+\zeta K/\rho)^2n^2}{(1+\chi_0 e^{-\zeta})^{5/2}(1+\zeta K/\rho)^3} e^{-\zeta} d\zeta;
\nonumber\\
&=& n_0 \chi_0 \frac{1}{4} \left[\frac12 \chi_0-2\frac{K}{\rho}-\frac{5}{12}\chi_0^2
-\frac32 \chi_0 \frac{K}{\rho}
+10 (\frac{K}{\rho})^2
\pm \cdots\right]
\label{eq.chi3}
\end{eqnarray}

For the fifth order in $\tan z_0$ up to second mixed order
\begin{eqnarray}
\gamma_5 
&=& \rho n_0 \frac{3}{16} \int_0^\infty \frac{(\rho^2n_0^2-(\rho+\zeta K)^2n^2)^2}{n^7(\rho+\zeta K)^5}\chi_0 e^{-\zeta} d\zeta
\nonumber\\
&=& n_0 \chi_0 \frac{3}{16} \int_0^\infty \frac{(n_0^2-(1+\zeta K/\rho)^2n^2)^2}{(1+\chi_0 e^{-\zeta})^{7/2}(1+\zeta K/\rho)^5} e^{-\zeta} d\zeta
\nonumber\\
&=& n_0 \chi_0 \frac{3}{16} \left[\frac13 \chi_0^2 -3\chi_0 \frac{K}{\rho}
+8 (\frac{K}{\rho})^2
\pm \cdots\right]
.
\label{eq.chi5}
\end{eqnarray}

More $\gamma_i$ coefficients are in the function \texttt{gamm()} in the file \texttt{RefrExpo.cxx}
in the source code (see App.\ \ref{app.prog}).

\begin{figure}
\includegraphics[scale=0.5]{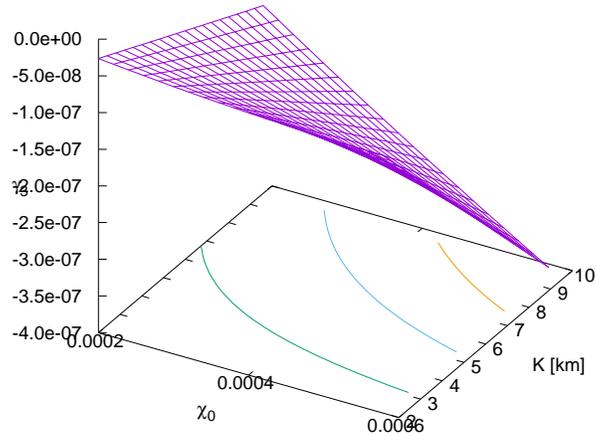}
\caption{The coefficient $\gamma_3$ in radians as a function of $\chi_0$ and $K$ at $\rho=6380$ km.
$2\times 10^{-7}$ rad are equivalent to 40 mas at $z_0=45^\circ$.}
\label{fig.R3}
\end{figure}

The difference between including only $\tan z_0$ and including all four terms
up to $\tan^7 z_0$ is
for example 0.8 arcsec (112.60 arcsec versus 111.79 arcsec) for 
$z_0=70^\circ$ at $\chi_0=4\times 10^{-4}$ and $K/\rho=9.6/6380$.
Then $\tan z_0=2.7$, and the dominating contribution to the 0.8 arcsec
is the cube of 2.7 multiplied by the characteristic $\gamma_3\approx 40$ mas
of Figure \ref{fig.R3}.

\begin{figure}
\includegraphics[scale=0.5]{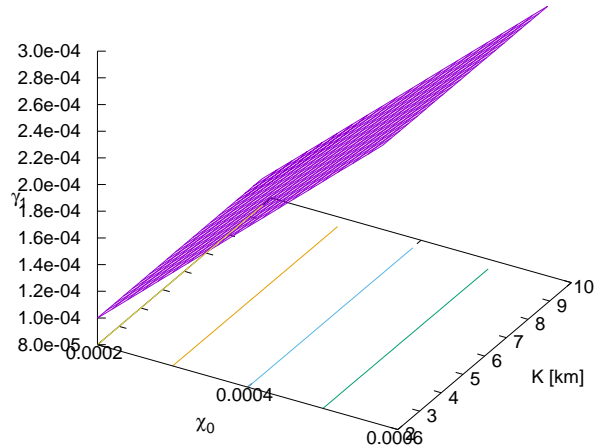}
\caption{The coefficient $\gamma_1$ in radians as a function of $\chi_0$ and $K$ at $\rho=6380$ km.
At this resolution there is almost no dependence on the scale height $K$, so $\gamma_1\approx n_0\chi_0/2$.
$2\times 10^{-4}$ rad is equivalent to 41 arcsec at $z_0=45^\circ$.
}
\label{fig.R1}
\end{figure}

\begin{figure}
\includegraphics[scale=0.5]{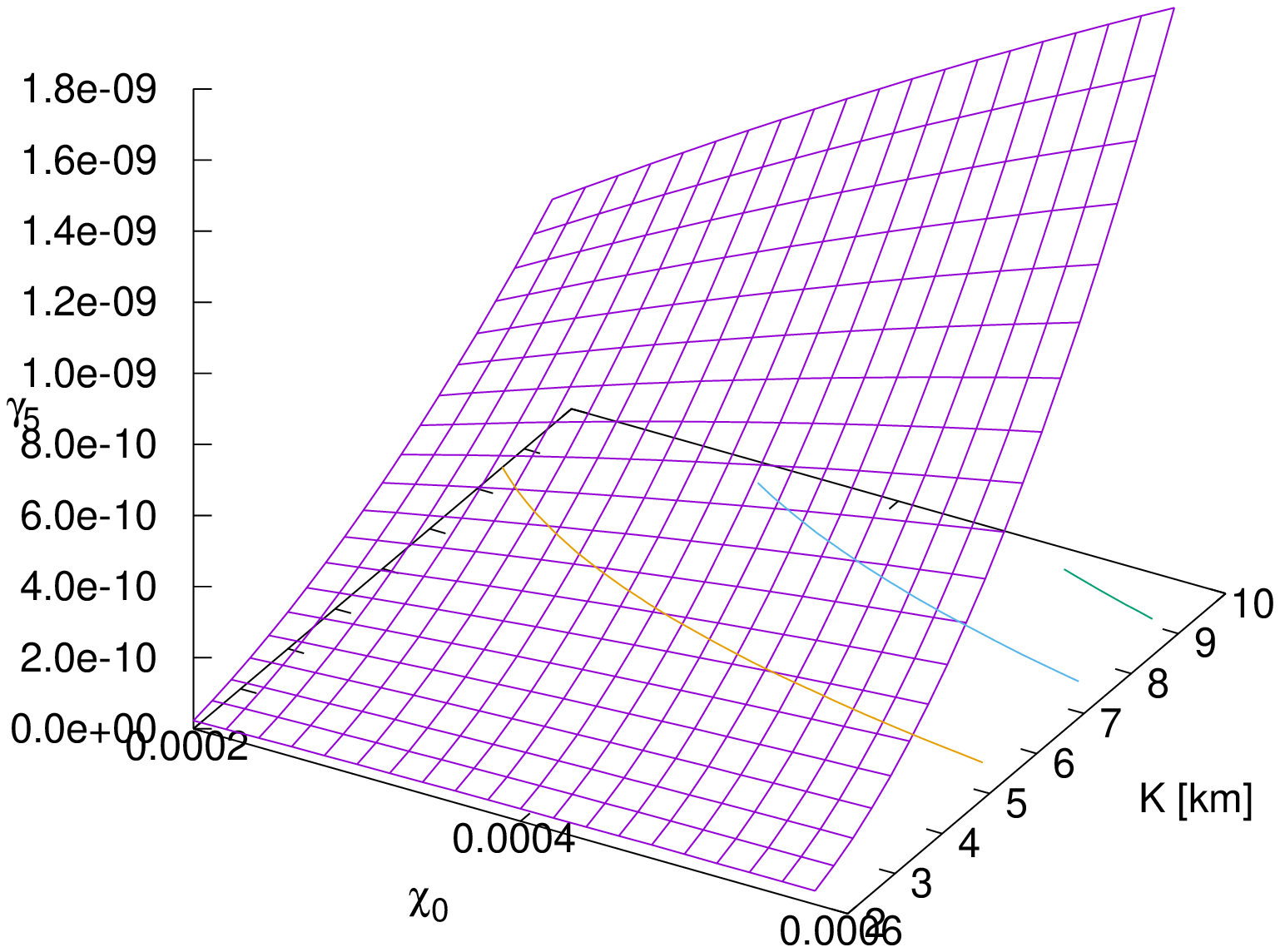}
\caption{The coefficient $\gamma_5$ in radians as a function of $\chi_0$ and $K$ at $\rho=6380$ km.
$10^{-9}$ rad are equivalent to 0.2 mas at $z_0=45^\circ$.}
\label{fig.R5}
\end{figure}

\section{Active Optics Aberrations}
The implication of the previous section for telescope operation is that
the telescope axis is corrected by a tilt $R$ while acquiring an object
at an astrometric zenith angle $z$. 
Which further imprint does the asymmetry
of the global spherical atmosphere leave if the telescope's primary mirror
collects light from a beam which does not have a perfect cylindrical symmetry
because the center of curvature of the atmosphere is not on the telescope axis?
Which quasi-static aberrations beyond
that type of `lowest Zernike order' appear, and how much is the wavefront
already curved as it hits the primary? 

We answer this question quantitatively by representing the entrance pupil of the primary
as a (continuous) set of baselines, with the same mathematical tools as for interferometric baselines \cite{MatharBA14}.

\subsection{Optical Path Length}
In a global coordinate system centered at the Earth center, the star light is represented 
by a bundle of parallel rays approaching the Earth starting at very large values on
the positive $x$-axis. (This does not restrict the results because we may tilt the
Earth axis here such that the star is in the equatorial plane.) The center of the primary mirror has a distance $\rho$
from the Earth center and is located at 
geodetic latitude $\varphi$,
\begin{equation}
\vec m(0,0) = \rho\left( \begin{array}{c} \cos \varphi \\ 0 \\ \sin\varphi \end{array}\right).
\end{equation}
The astrometric zenith angle $z$ is $z = \varphi$, representing a direction $(1,0,0)$
in the global coordinate system. The actual pointing axis of the telescope
is along the zenith angle $z_0=\varphi -R$, which is the direction $(\cos R,0, \sin R)$
in the global coordinate system. The points in the entrance pupil of the primary mirror
have local coordinates $m_{h,v}$ where $m_h$ points along the local horizontal and $m_v$
from the center to the upper rim, and where $m_h^2+m_v^2$ is smaller than the squared radius of the primary.

The task is to calculate the optical path length
\begin{equation}
L= \int_{n=1}^{n=n_0} n \frac{dr}{\cos \psi},
\label{eq.Lairm}
\end{equation}
the air mass weighted with the refractive index, given $m_h$ and $m_v$ in the entrance pupil,
or, more precisely, recovering the differential optical path length relative to the 
center. This becomes a mixed boundary problem: we define locations of
the rays in the entrance pupil, but directions (their tilts, derivatives) 
at some common virtual entrance pupil high above the atmosphere.

\subsection{Spanning the Entrance Pupil}
\begin{figure}
\includegraphics[scale=0.4]{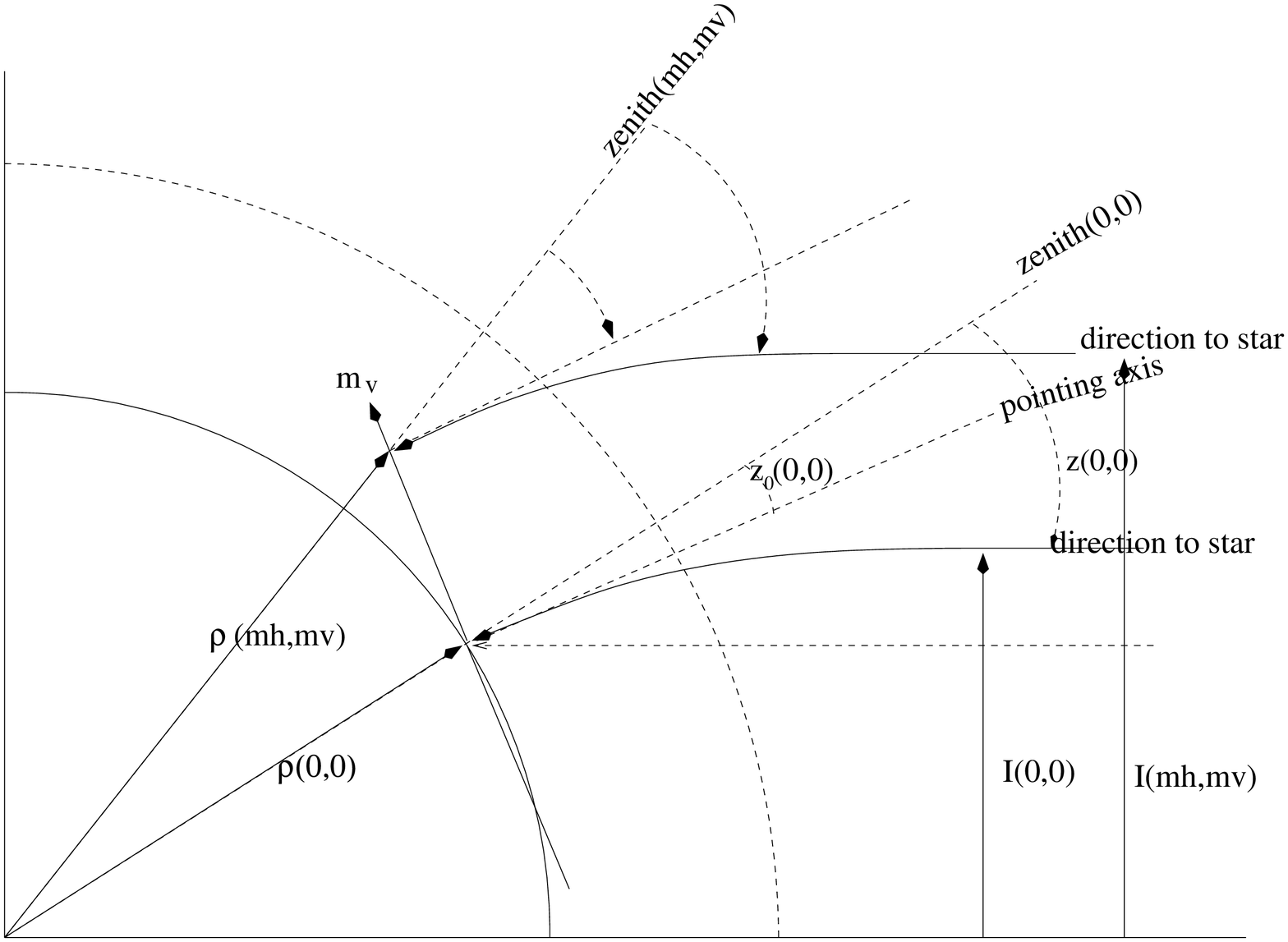}
\caption{The telescope points into a direction described by $z_0(0,0)$. $m_h$ (not shown, pointing
into the drawing plane)
and $m_v$ are directions  orthogonal to that pointing direction. Two bent rays from a common star
run parallel high above the atmosphere and reach the pupil with two different sets
of zenith angles $z$, $z_0$, impact parameters $I$, and air densities.
(Not to scale.)}
\label{fig.mcoo}
\end{figure}
The global position of a point in the circular entrance pupil is
the center of the pupil plus two components perpendicular to the pointing direction $z_0(0,0)$
with local horizontal and vertical Cartesian coordinates $m_h$ and $m_v$:
\begin{equation}
\vec m(m_h,m_v) = \vec m(0,0) + m_v\left(\begin{array}{c}-\sin R(0,0) \\ 0 \\ \cos R(0,0)\end{array}\right) 
+m_h \left(\begin{array}{c}0\\ 1 \\ 0\end{array}\right)
=
\left(\begin{array}{c}\rho(0,0)\cos z(0,0)-m_v\sin R(0,0) \\ m_h \\ \rho(0,0)\sin z(0,0) +m_v \cos R(0,0)\end{array}\right).
\label{eq.mofhv}
\end{equation}
The zenith angles $z$ and $z_0$, the refractive indices $n$ and $\chi_0$, and the distances $\rho$ and $r$
to the Earth center are functions of the $(m_h,m_v)$ coordinates of the pupil coordinates.
The refinement of Figure \ref{fig.sph} for a bundle of rays which enter from the
right leads to Figure \ref{fig.mcoo}.
The incidence plane that contains the refracted ray along its entire path is tilted by the angle $\phi$
out of the drawing plane of Figure \ref{fig.mcoo}; the ratio of
the lower two components of Eq.\ (\ref{eq.mofhv})  yields
\begin{equation}
\tan \phi = \frac{m_h}{\rho(0,0)\sin z(0,0)+m_v\cos R(0,0)}.
\label{eq.iplane}
\end{equation}

The distance of such a point to the Earth  center is $|\vec m|$ and defines the variation
of the local refractive index due to the variation of the air density in the entrance pupil:
\begin{equation}
|\vec m(m_h,m_v)| = \rho(m_h,m_v) = \sqrt{\rho^2(0,0)+m_h^2+m_v^2+2\rho(0,0) m_v \sin z_0(0,0) }.
\end{equation}
The direction of the zenith of these points in the entrance pupil
is $\vec m/|\vec m|$, defined as to split the entrance pupil into
smaller areas which share a common pointing axis but which collect
light that is bent along individual curved paths.
The astrometric zenith angles $z(m_h,m_v)$ of points in  the entrance pupil
are aligned (above the atmosphere) to a common direction; by building
the dot product of the unit vector $\vec m/|\vec m|$ with the vector (1,0,0)
of the incoming light,
$z$ is fixed by
\begin{eqnarray}
\cos z(m_h,m_v) &=& \frac{1}{|\vec m(m_h,m_v)|} [\rho(0,0) \cos z(0,0) - m_v\sin R(0,0)]
\label{eq.cosz}
\\
&=& \frac{\rho(0,0) \cos z(0,0) - m_v\sin R(0,0)}
{
\sqrt{\rho^2(0,0)+m_h^2+m_v^2+2\rho(0,0) m_v \sin z_0(0,0) }
}
.
\end{eqnarray}
This is inconvenient because the application further down is formulated in
terms of the true zenith angle $z_0(m_h,m_v)$.
To obtain $z_0(m_h,m_v)$ we may
\begin{itemize}
\item
recompute $R$ in terms of $\tan z$ \cite{MatharBA14}: Define coefficients $\bar\gamma_i$ via
\begin{equation}
R\equiv \sum_{i=1,3,5,7,\ldots} \bar \gamma_i \tan^i z.
\label{eq.gammbdef}
\end{equation}
There is a Taylor expansion of $\tan z_0$ in orders of $R$, then $R$
can be rephrased with Eq. (\ref{eq.gammbdef}):
\begin{eqnarray}
\tan z_0&=& \tan(z-R) = \tan z -(1+\tan^2 z)R+\tan(z)[1+\tan^2 z]R^2-[\frac43 \tan^2z +\tan^4 z+\frac12]R^3+O(R^4)
\nonumber\\
&=& (1-\bar \gamma_1)\tan z -(\bar \gamma_1-\bar \gamma^2_1 +\frac13 \bar \gamma^3_1+\bar \gamma_3) \tan^3z + O(\tan^5z).
\end{eqnarray}
Inserting odd powers of this into (\ref{eq.gammidef}) yields
\begin{equation}
\gamma_1 \tan z_0+\gamma_3 \tan^3z_0+O(\tan^5z_0)
=
\gamma_1(1-\bar \gamma_1)\tan z + [\gamma_1(-\bar \gamma_1-\bar\gamma_3-\frac13 \bar\gamma_1^3+\bar\gamma_1^2)
+\gamma_3(1-\bar\gamma_1)^3]\tan^3z +O(\tan^5z).
\end{equation}
Equating this with (\ref{eq.gammbdef}) and comparison of equal powers of $\tan z$ on both sides
gives a system of self-consistent equations
\begin{eqnarray}
\bar \gamma_1 &=& \gamma_1(1-\bar\gamma_1),\\
\bar \gamma_3 &=& \gamma_1(-\bar\gamma_1-\bar\gamma_3-\frac13 \bar\gamma_1^3+\bar\gamma_1^2)+\gamma_3(1-\bar\gamma_1)^3,
\end{eqnarray}
which are  solved in that order by
\begin{eqnarray}
\bar \gamma_1&=& \frac{\gamma_1}{1+\gamma_1},\\
\bar \gamma_3&=& - \frac{\gamma_1^4+3\gamma_1^3+3\gamma_1^2-3\gamma_3}{3(1+\gamma_1)^4}.
\end{eqnarray}
More terms are listed in the function \texttt{gammBar()} in the 
file \texttt{RefrExpo.cxx} in the source code.
Because the number of terms in these relations grows rapidly as a function
of the index $i$, this is numerically less stable and not used here.
\item
or solve (\ref{eq.R}),
$z-z_0=\sum_{i=1,3,5,\ldots} \gamma_i \tan^i z_0$, for $z_0$ via a few iterations of a Newton method:
\begin{equation}
z_0^{(j)} = z_0^{(j-1)}-\frac{z_0-z+\sum_{i=1,3,5,\ldots} \gamma_i \tan^i z_0^{(j-1)}}{1+\sum_{i=1,3,5,\ldots}i\gamma_i \tan^{(i-1)} z_0^{(j-1)}[1+\tan^2 z_0^{(j-1)}] },
\end{equation}
starting with a first estimate $z_0^{(0)}=z$.
\end{itemize}

Each point in the entrance pupil has a path length $L(m_h,m_v)$ of a ray that terminates there. Rephrasing (\ref{eq.Lairm}) with the 
Fresnel law (see \cite{MatharBA14}) gives
\begin{equation}
L = \int_{n=1}^{n=n_0} rn^2 \frac{dr}{\sqrt{n^2r^2-n_0^2\rho^2\sin^2 z_0}}.
\end{equation}
The aberration across the pupil, measured in length units, 
is the path length difference between a ray that ends at a general general point
and a ray that ends at the center:
\begin{eqnarray}
\Delta L 
= 
\int_{n=1}^{n=n_0(m_h,m_v)} rn^2 \frac{dr}{\sqrt{n^2r^2-n_0(m_h,m_v)^2\rho(m_h,m_v)^2\sin^2 z_0(m_h,m_v)}}
\nonumber
\\
-
\int_{n=1}^{n=n_0(0,0)} rn^2 \frac{dr}{\sqrt{n^2r^2-n_0(0,0)^2\rho(0,0)^2\sin^2 z_0(0,0)}}
.
\end{eqnarray}

\subsection{Series Expansions}
The first integral may be split by prolonging the ray to the atmospheric layer
of the center and by backing up from there to the layer in the pupil:
\begin{eqnarray}
\Delta L 
= 
\int_{n=1}^{n=n_0(0,0)} rn^2 \frac{dr}{\sqrt{n^2r^2-n_0(m_h,m_v)^2\rho(m_h,m_v)^2\sin^2 z_0(m_h,m_v)}}
\nonumber
\\
+\int_{n=n_0(0,0)}^{n=n_0(m_h,m_v)} rn^2 \frac{dr}{\sqrt{n^2r^2-n_0(m_h,m_v)^2\rho(m_h,m_v)^2\sin^2z_0(m_h,m_v)}}
\nonumber
\\
-
\int_{n=1}^{n=n_0(0,0)} rn^2 \frac{dr}{\sqrt{n^2r^2-n_0(0,0)^2\rho(0,0)^2\sin^2z_0(0,0)}}
.
\label{eq.Lsplit}
\end{eqnarray}
Define impact parameters $I$
\begin{equation}
I(m_h,m_v)\equiv n_0(m_h,m_v)\rho(m_h,m_v)\sin z_0(m_h,m_v)
.
\end{equation}

The aberration is split into a contribution of the second integral and a compound term
of the first and third integral, $\Delta L=\Delta L^{(1)} + \Delta L^{(2)}$.
The second integral is rephrased as a function of the variables of Section \ref{sec.expo}:
\begin{eqnarray}
\Delta L^{(1)}
=
\int_{n=n_0(0,0)}^{n=n_0(m_h,m_v)} rn^2 \frac{dr}{\sqrt{n^2r^2-I^2(m_h,m_v)}}
=
\int_{r=\rho(0,0)}^{r=\rho(m_h,m_v)} r(1+\chi) \frac{dr}{\sqrt{(1+\chi)r^2-I^2(m_h,m_v)}}
\nonumber \\
=
K\int_{\zeta=0}^{\zeta=\zeta(m_h,m_v)} (1+\zeta K/\rho)(1+\chi_0e^{-\zeta}) \frac{d\zeta}{\sqrt{(1+\chi_0e^{-\zeta})(1+\zeta K/\rho)^2
-I^2(m_h,m_v)/\rho^2}}
.
\end{eqnarray}
This is expanded in a bivariate power  series for small $\chi_0$ and $K/\rho$:
\begin{eqnarray}
\Delta L^{(1)}
=
K\int_{\zeta=0}^{\zeta=\zeta(m_h,m_v)} 
\big[
\frac{1}{(1-\hat I^2)^{1/2}}
-\frac{\hat I^2}{(1-\hat I^2)^{3/2}}\zeta \frac{K}{\rho}
+\frac12 \frac{1-2\hat I^2}{(1-\hat I^2)^{3/2}} e^{-\zeta} \chi_0
+\frac32 \frac{\hat I^2}{(1-\hat I^2)^{5/2}}\zeta^2 (\frac{K}{\rho})^2
\nonumber \\
+\frac12 \frac{\hat I^2(1+2\hat I^2)}{(1-\hat I^2)^{5/2}}\zeta e^{-\zeta} \frac{K}{\rho}\chi_0
-\frac18 \frac{\hat 1-4\hat I^2}{(1-\hat I^2)^{5/2}}e^{-2\zeta} \chi_0^2
\nonumber \\
-\frac12 \frac{\hat I^2(4+\hat I^2)}{(1-\hat I^2)^{7/2}}\zeta^3 (\frac{K}{\rho})^3
-\frac34 \frac{\hat I^2(1+4\hat I^2)}{(1-\hat I^2)^{7/2}}\zeta^2 e^{-\zeta} (\frac{K}{\rho})^2 \chi_0
\nonumber \\
-\frac38 \frac{\hat I^2(1+4\hat I^2)}{(1-\hat I^2)^{7/2}}\zeta e^{-2\zeta} \frac{K}{\rho} \chi_0^2
+\frac{1}{16} \frac{1-6\hat I^2}{(1-\hat I^2)^{7/2}} e^{-3\zeta} \chi_0^3
+\cdots
\big]d\zeta
\end{eqnarray}
where $\chi_0=\chi_0(0,0)$, $\rho=\rho(0,0)$ and $\hat I\equiv I(m_h,m_v)/\rho$.
Term-by-term integration yields
\begin{eqnarray}
\Delta L^{(1)}
=
K
\big[
\frac{1}{(1-\hat I^2)^{1/2}}\zeta
-\frac12 \frac{\hat I^2}{(1-\hat I^2)^{3/2}}\zeta^2 \frac{K}{\rho}
-\frac12 \frac{1-2\hat I^2}{(1-\hat I^2)^{3/2}} e^{-\zeta} \chi_0
+\frac12 \frac{\hat I^2}{(1-\hat I^2)^{5/2}}\zeta^3 (\frac{K}{\rho})^2
\nonumber \\
-\frac12 \frac{\hat I^2(1+2\hat I^2)}{(1-\hat I^2)^{5/2}}(1+\zeta) e^{-\zeta} \frac{K}{\rho}\chi_0
+\frac{1}{16} \frac{\hat 1-4\hat I^2}{(1-\hat I^2)^{5/2}}e^{-2\zeta} \chi_0^2
\nonumber \\
-\frac18 \frac{\hat I^2(4+\hat I^2)}{(1-\hat I^2)^{7/2}}\zeta^4 (\frac{K}{\rho})^3
+\frac34 \frac{\hat I^2(1+4\hat I^2)}{(1-\hat I^2)^{7/2}}(2+2\zeta+\zeta^2) e^{-\zeta} (\frac{K}{\rho})^2 \chi_0
\nonumber \\
+\frac{3}{32} \frac{\hat I^2(1+4\hat I^2)}{(1-\hat I^2)^{7/2}}(1+2\zeta) e^{-2\zeta} \frac{K}{\rho} \chi_0^2
-\frac{1}{48} \frac{1-6\hat I^2}{(1-\hat I^2)^{7/2}} e^{-3\zeta} \chi_0^3
+\cdots
\big]_{\zeta=0}^{\zeta(m_h,m_v)}
.
\label{eq.dl1}
\end{eqnarray}
More terms are listed in the function \texttt{deltaL1Int()} in the file \texttt{M1aberr.cxx} in the source code.

The first and third term of (\ref{eq.Lsplit}) are expanded in powers of $\Delta I\equiv I(0,0)-I(m_h,m_v)$,
which is of the order of the pupil radius because the atmosphere is thin:
\begin{eqnarray}
\frac{1}{\sqrt{n^2r^2-I^2(m_h,m_v)}}-\frac{1}{\sqrt{n^2r^2-I^2(0,0)}}
=
-\frac{I(m_h,m_v)}{(n^2r^2-I(m_h,m_v))^{3/2}} \Delta I
\nonumber \\
-\frac12 \frac{n^2r^2+2I^2(m_h,m_v)}{(n^2r^2-I(m_h,m_v))^{5/2}} (\Delta I)^2
\nonumber \\
-\frac32 \frac{[3n^2r^2+2I^2(m_h,m_v)]I(m_h,m_v)}{(n^2r^2-I(m_h,m_v))^{7/2}} (\Delta I)^3
\nonumber \\
-\frac18 \frac{3n^4r^4+24n^2r^2I^2(m_h,m_v)+8I^4(m_h,m_v)}{(n^2r^2-I(m_h,m_v))^{9/2}} (\Delta I)^4
+\cdots
.
\end{eqnarray}
\begin{eqnarray}
\Delta L^{(2)} & = & \int_{r=\infty}^{\rho(0,0)}dr\left[\frac{rn^2}{\sqrt{n^2r^2-I^2(m_h,m_v)}}-\frac{rn^2}{\sqrt{n^2r^2-I^2(0,0)}}\right]
\nonumber \\
&= &
K\int_{\zeta=\infty}^0
d\zeta
\big [
-\frac{(1+\zeta K/\rho)(1+\chi_0e^{-\zeta}) \hat I}{(n^2r^2-\hat I)^{3/2}} (\Delta I)/\rho
\nonumber \\
&&
-\frac12 \frac{(1+\zeta K/\rho)n^2[n^2(1+\zeta K/\rho)^2+2\hat I^2]}{(n^2r^2-\hat I)^{5/2}} (\Delta I/\rho)^2
\nonumber \\
&&
-\frac32 \frac{(1+\zeta K/\rho)n^2[3n^2(1+\zeta K/\rho)^2+2\hat I^2]\hat I}{(n^2r^2-\hat I)^{7/2}} (\Delta I/\rho)^3
\nonumber \\
&&
-\frac18 \frac{(1+\zeta K/\rho)n^2[3n^4(1+\zeta K/\rho)^4+24n^2(1+\zeta K/\rho)^2\hat I^2+8\hat I^4]}{(n^2r^2-\hat I)^{9/2}} (\Delta I/\rho)^4
+\cdots
\big]
.
\end{eqnarray}

In a double expansion in orders of $K/\rho$ and $\chi_0$, we integrate the orders of $\chi_0^0$ all in once:
\begin{eqnarray}
\Delta L^{(2)} \stackrel{\chi_0\to 0, n\to 1}{\longrightarrow} && \int_{r=\infty}^{\rho(0,0)}
dr \big[\frac{r}{\sqrt{r^2-I^2(m_h,m_v)}}-\frac{r}{\sqrt{r^2-I^2(0,0)}}\big]
\nonumber \\
&&
=\sqrt{\rho^2-I^2(m_h,m_v)}-\sqrt{\rho^2-I^2(0,0)}.
\end{eqnarray}
The full expansion in powers of $K/\rho$, $\chi_0$ and $\Delta I/\rho$ can then be written as
\begin{eqnarray}
\Delta L^{(2)} 
=\sqrt{\rho^2-I^2(m_h,m_v)}-\sqrt{\rho^2-I^2(0,0)} + K \sum_{k\ge 0} \sum_{c\ge 1} \sum_{i\ge 1} \alpha_{k,c,i}
(K/\rho)^k \chi_0^c (\Delta I/\rho)^i\frac{1}{(1-\hat I^2)^{1/2+k+c+i}}
.
\label{eq.dl2}
\end{eqnarray}
where $K/\rho$ is of the order of 10 kilometers over the Earth radius,
$\chi_0$ is of the order of a few $\times 10^{-4}$ for typical optical telescopes,
and where $\Delta I/\rho$ is of the order of the primary mirror radius over the Earth radius.
Some $\alpha_{k,c,i}$ are tabulated in Table \ref{tab.alpha}; an extended
set is in the function \texttt{deltaL2Series()} in the file \texttt{M1aberr.cxx} of the source code.

\begin{table}
\begin{tabular}{rrr|l}
$k$ & $c$ & $i$ & $\alpha$\\
\hline
0 & 1 & 1 & $ -\frac{\hat I(1+2\hat I^2)}{2}$\\
0 & 1 & 2 & $-\frac{1+10\hat I^2+4\hat I^4}{4}$\\
0 & 1 & 3 & $-\frac{(9+22\hat I^2+4\hat I^4)\hat I}{4}$\\
1 & 1 & 1 & $ \frac{(2+11\hat I^2+2\hat I^4)\hat I}{2}$\\
1 & 1 & 2 & $ \frac{2+45\hat I^2+54\hat I^4+4\hat I^6}{4}$\\
1 & 1 & 3 & $ \frac{(36+177\hat I^2+98\hat I^4+4\hat I^6)\hat I}{4}$\\
0 & 2 & 1 & $ \frac{3(1+4\hat I^2)\hat I}{16}$\\
0 & 2 & 2 & $ \frac{3+54\hat I^2+48 \hat I^4}{32}$\\
0 & 2 & 3 & $ \frac{5(9+38\hat I^2+12 \hat I^4)\hat I}{32}$\\
2 & 1 & 1 & $ -\frac{3(2+21\hat I^2+12 \hat I^4)\hat I}{2}$\\
\end{tabular}
\caption{$\alpha$-coefficients of Eq. (\ref{eq.dl2}) for small powers of the three expansion coefficients.}
\label{tab.alpha}
\end{table}

\subsection{Numerical Example}
\begin{figure}
\includegraphics[scale=0.6]{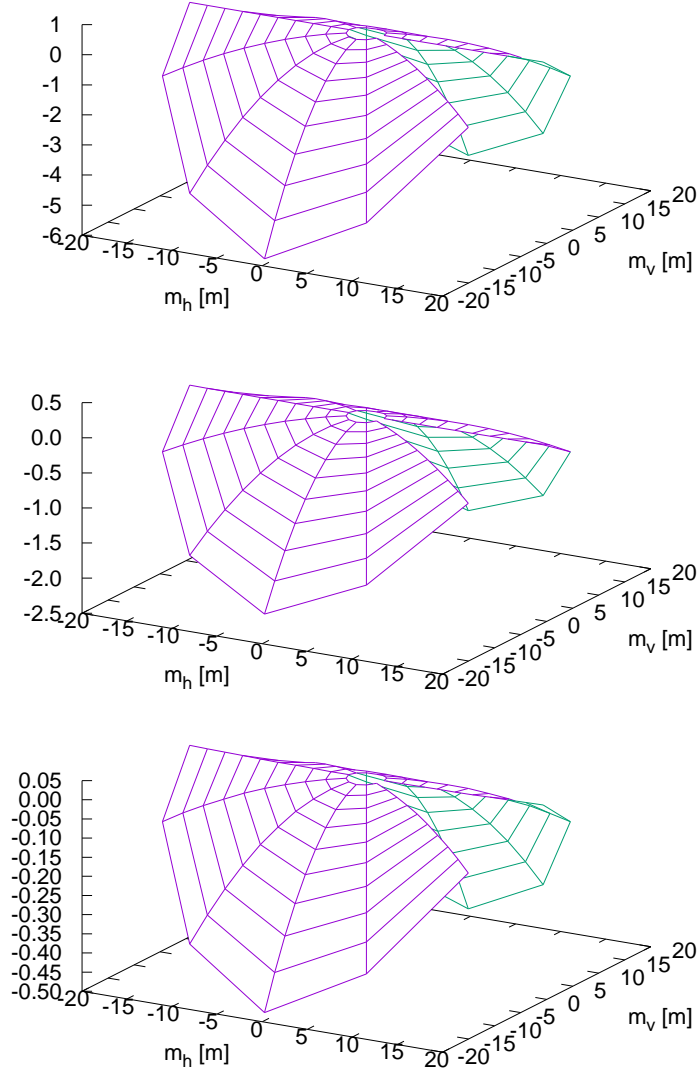}
\caption{The optical path length difference $\Delta L$ in micrometer
in the entrance pupil of a 40-m telescope
for zenith angles of 60 (top), 40 (middle) and 20 degrees (bottom) at $\chi_0=3.9\times 10^{-4}$.}
\label{fig.ELT}
\end{figure}

The result for prototypical E-ELT variables 
is shown in Figure \ref{fig.ELT}:
With Ciddor's formulas for the index of refraction we assume a wavelength of 500 nm,
a temperature of 10$^\circ$ C, a pressure of 697 hPa, a CO$_2$ content of 470 ppm,
so the refractive index is 1.000195229, and  $\chi_0=3.9\times 10^{-4}$.
The semi-major axis and flattening of the WGS84  \cite{NIMA8350}
at a geographic latitude of $\varphi=24^\circ 35'$,  
and sea level altitude 3046 m gives a distance $\rho(0,0)=6377.5$ km to the Earth center.
Results are plotted for a radius up to $\sqrt{m_h^2+m_v^2}\le 19.6$ m and a scale height of $K=9600$ m.
The conclusion is:
\begin{itemize}
\item
There is almost no residual optical path length along the horizontal ($m_h$)
axis (tip). 
\item
There is essentially no tilt component left, which means a tilt
of the telescope axis by $R$ as computed in Section \ref{sec.expo} removes the first-order Zernike terms,
as expected.
\item
The  residual difference of the path length of the upper and lower rim ($m_v$) relative
to the center is up to 6 $\mu$m of uni-axial defocus at a zenith distance of 60$^\circ$.
\item
The residual difference is approximately proportional to the
 residual air mass: $|\Delta L|\propto \frac{1}{\cos z}-1$.
\end{itemize}

\begin{figure}
\includegraphics[scale=0.6]{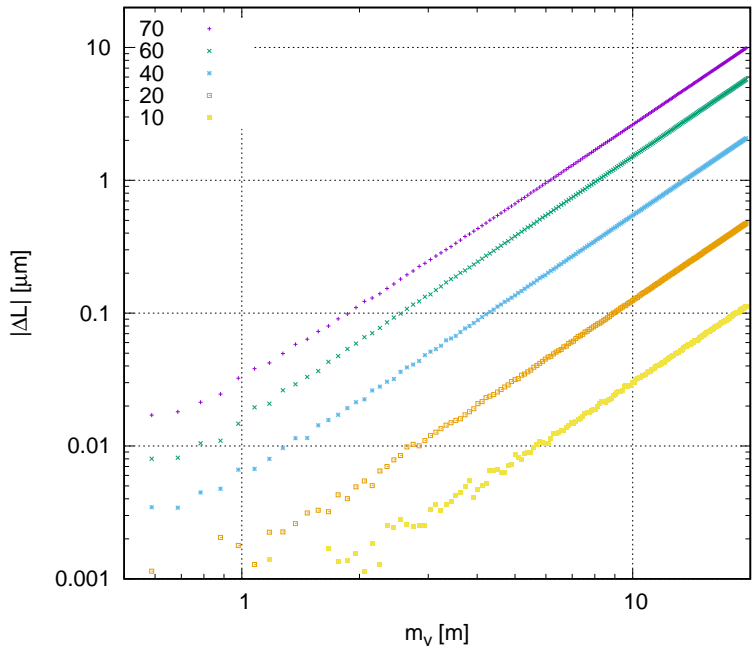}
\caption{The absolute value of optical path length difference $\Delta L$ in microns
in the entrance pupil between the upper rim and the center
for zenith angles of 70, 60, 40, 20 and 10 degrees (top to bottom). These are cuts through graphs
of Figure \ref{fig.ELT} at $m_h=0$.}
\label{fig.Raberr}
\end{figure}

We may zoom into such a graph and plot the absolute value of $\Delta L$ in a cut
from the center of the pupil to the upper rim, which yields Figure \ref{fig.Raberr}.
This plot with a double logarithmic scale of the optical path difference shows that 
\begin{itemize}
\item
the functions are almost straight lines for zenith distances up to 60$^\circ$; 
the aberration is approximately proportional to the squared distance (not a pure second-order Zernike term).
For a 8m-telescope at $z=60^\circ$ the path difference is only 0.26 $\mu$m.
\item
the numerical implementation has jitters on path length scales of typically 3 nanonmeters.
This is likely a consequence of rounding errors with the double-precision
arithmetics in the C++ source code, plus precision limits caused by cut-offs in the number
of taylor terms included while computing $\Delta L^{(1)}$ and 
$\Delta L^{(2)}$.
\end{itemize}

In a general telescope operation effects of the order calculated here
are probably not observed, because a focusing model that includes
flexure compensations or wind loads as a function of zenith angle
must handle basically the same data.

\begin{figure}
\includegraphics[scale=0.6]{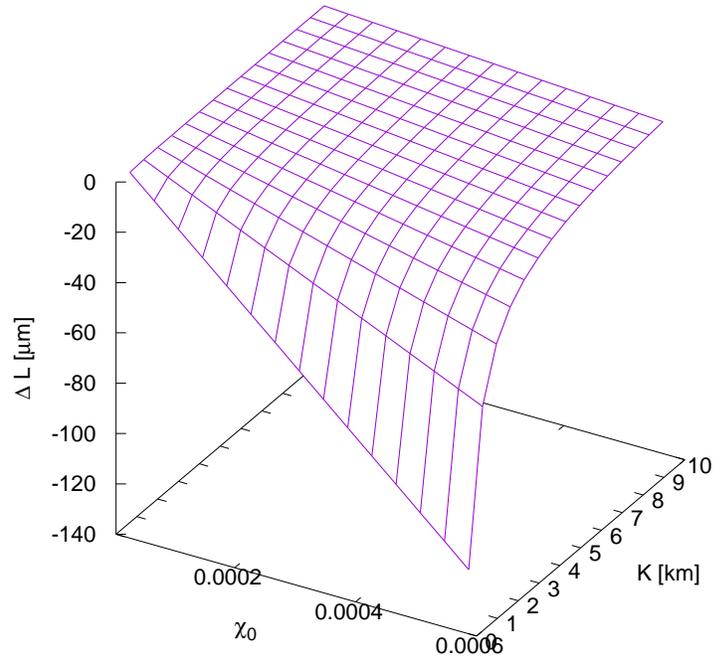}
\caption{The optical path length difference $\Delta L$ in microns
in the entrance pupil at the upper rim 19.6m away from the center
for a zenith angle of 60$^\circ$.
}
\label{fig.Raberr60d}
\end{figure}

With another cut through the parameter set we may keep $z_0$ and $m_v$ fixed and look at
$\Delta L$ as a function of the refractive index at the ground; this is relevant if
the adaptive optics system and the science cameras observe at different wavelengths and hence at different
$\chi_0$. The numerical data of Figure \ref{fig.Raberr60d} reveal that the change in
$\Delta L$ is practically proportional to $\chi_0$ as long as the scale height is fixed.
In consequence one may map relative differential chromatic differences in the
susceptibilities directly to relative differential aberrations in the entrance pupil.

\section{Summary}
In a barometric exponential model of the the air susceptibility profile
above the telescope we presented approximate analytical formulas 
\begin{enumerate}
\item
of the differential zenit angle $R$  as a power series of the tangent of the zenith
angle with expansion coefficients summarized in (\ref{eq.chi1}), (\ref{eq.chi3}) and (\ref{eq.chi5});
\item
of the path length distribution $\Delta L$ of rays in the entrance pupil
with two contributions summarized by (\ref{eq.dl1}) and (\ref{eq.dl2}).
\end{enumerate}

\clearpage
\appendix
\section{Segregated Scale Heights}\label{app.segr}
If gas mixtures of the air are treated as independent linear responses such that $\chi_0$ is 
an additive composition proportional to the weighted contribution of the air components, Equation (\ref{eq.chi1}) breaks
already in the first three terms in the square brackets if the scale heights of the molecular constituents differ.
The typical case is an additive mixture of dry air and water vapor, where dry air has a
scale height $K$ near 9 kilometers, but water vapor disappears 
at a scale height of approximately 2 kilometers \cite{TurnerGRL28,CahenJAM21,ParameswaranJAM29,OtarolaPASP122}.
To illustrate that variant, one would introduce a susceptibility with two scale heights $K_1$ and $K_2$,
\begin{equation}
\chi(r) = \chi_{01}e^{-(r-\rho)/K_1} + \chi_{02} e^{-(r-\rho)/K_2}
 = \chi_{01}e^{-(r-\rho)/K_1} + (1 \leftrightarrow 2),
\end{equation}
where $1\leftrightarrow 2$ means the previous terms is repeated where the roles of $K_1$ and $K_2$ are swapped,
and also the roles of $\chi_{01}$ and $\chi_{02}$ are swapped.
The lowest order expansion 
for small $\hat K_1\equiv K_1/\rho$ and $\hat K_2\equiv K_2/\rho$ and small $\chi_{01}$, $\chi_{02}$ is
\begin{eqnarray}
\gamma_1 
&=& \frac12 \rho n_0\int_1^{n_0} \frac{1}{n^3r}d\chi
\nonumber \\
&=& \frac12 \rho n_0 \int_{\rho} ^\infty \frac{1}{[1+\chi_{01}e^{-(r-\rho)/K_1} + \chi_{02} e^{-(r-\rho)/K_2}]^{3/2}r}
     [
     \frac{1}{K_1}\chi_{01}e^{-(r-\rho)/K_1}
     +\frac{1}{K_2}\chi_{02}e^{-(r-\rho)/K_2}
     ] dr
\nonumber \\
&=& \frac12 \rho n_0 \int_{0} ^\infty \frac{1}{[1+\chi_{01}e^{-r/K_1} + \chi_{02} e^{-r/K_2}]^{3/2}(r+\rho)}
     [
     \frac{1}{K_1}\chi_{01}e^{-r/K_1}
     +\frac{1}{K_2}\chi_{02}e^{-r/K_2}
     ] dr
\nonumber \\
\nonumber \\
&=& \frac12 \frac{\rho}{K_1} n_0\chi_{01} \int_{0} ^\infty \frac{e^{-r/K_1}}{[1+\chi_{01}e^{-r/K_1} + \chi_{02} e^{-r/K_2}]^{3/2}(r+\rho)}
     dr
     +(1\leftrightarrow 2)
\nonumber \\
&\approx & 
     \frac{\chi_{01}}{8(\hat K_1 +\hat K_2)}
     \left[4(\hat K_1+\hat K_2-\hat K_1\hat K_2-\hat K_1^2) +2\chi_{02} \hat K_1-4\chi_{02}\hat K_2-\chi_{01} (\hat K_1+\hat K_2)
     \right]
     +(1\leftrightarrow 2)
.
\end{eqnarray}

\section{Cassini's model}
\subsection{Tilt Correction}

The Cassini model of a homogeneous single-layer atmosphere of height $h$ leads to a difference
between observed and astrometric zenith angles of
\cite[\S 9.5]{YoungAJ127}\cite{CorbardMNRAS483,vandenBornMNRAS}
\begin{equation}
R^{(C)} =  \arcsin\left(\frac{n_0\rho\sin z_0}{\rho+h}\right)-\arcsin\left(\frac{\rho\sin z_0}{\rho+h}\right)
.
\label{eq.Rcas}
\end{equation}

\begin{figure}
\includegraphics[scale=0.4]{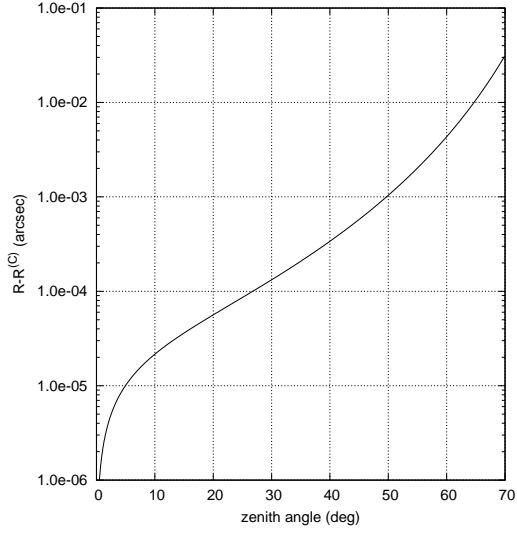}
\caption{The difference between the refraction predicted by (\ref{eq.Rcas}) and the model of exponential slope, 
$R-R^{(C)}$, as a function of zenith distance with $\chi_0=0.000568081$
and $\rho=6377.36$ km and $K=9.6$ km. Less than 30 mas for zenith angles below $70^\circ$.
}
\label{fig.Rcas}
\end{figure}

The bivariate Taylor expansion in orders of $n_0-1$ and $q$ starts
\begin{eqnarray}
R^{(C)}&=& (n_0-1)\tan z_0\big[
1
-\frac{q}{\cos^2 z_0}
+\frac12 (n_0-1)\tan^2 z_0
+\frac12 \frac{q^2(3-\cos^2 z_0)}{\cos^4 z_0}
-\frac32 \frac{(n_0-1)q\tan^2 z_0}{\cos^2 z_0}
\nonumber \\
&&\quad
+\frac16 \frac{(n_0-1)^2\tan^2z_0(3-2\cos^2z_0)}{\cos^2 z_0}
+\cdots
\big]
\end{eqnarray}
where $q=h/\rho$. The small difference $R-R^{(C)}$ for a parameter set of $\rho=6377.36$ km, $h=K=9.6$ km
and $n_0=1.000284$ is illustrated in Figure \ref{fig.Rcas}.

Expansion of $R^{(C)}$ in powers of $\tan z_0$ defines coefficients $\gamma_i^{(C)}$:
\begin{equation}
R^{(C)}\equiv \sum_{i=1,3,5,\ldots} \gamma_i^{(C)} \tan^iz_0,
\end{equation}
such that up to third mixed order in $\chi_0$ and $q$
\begin{equation}
\gamma_1^{(C)} = 
n_0\chi_0\frac12\left[1-\frac34 \chi_0 -q +\frac{5}{8}\chi_0^2 +\frac34 \chi_0 q+q^2+\ldots\right],
\label{eq.Cchi1}
\end{equation}
\begin{equation}
\gamma_3^{(C)} = 
n_0\chi_0\frac14\left[\frac12 \chi_0-2q -\frac{5}{12}\chi_0^2 +5q^2+\ldots\right],
\label{eq.Cchi3}
\end{equation}
\begin{equation}
\gamma_5^{(C)} = 
n_0\chi_0\frac{3}{16}\left[\frac13 \chi_0^2 -2\chi_0q +4q^2+\ldots\right].
\end{equation}
The difference between the Cassini model and the exponential model is given 
by subtracting (\ref{eq.Cchi1}) and (\ref{eq.Cchi3}) from (\ref{eq.chi1}) and (\ref{eq.chi3}),
\begin{equation}
R-R^{(c)} \approx \frac12 n_0\chi_0 
\left(-\frac34 \chi_0 q+q^2+\cdots\right)
\tan z_0 
+\frac14 n_0\chi_0 
\left(-\frac32 \chi_0 q+5q^2+\cdots\right)
\tan^3 z_0 
+\cdots
\end{equation}

\subsection{Aberration}
\begin{figure}
\includegraphics[scale=0.6]{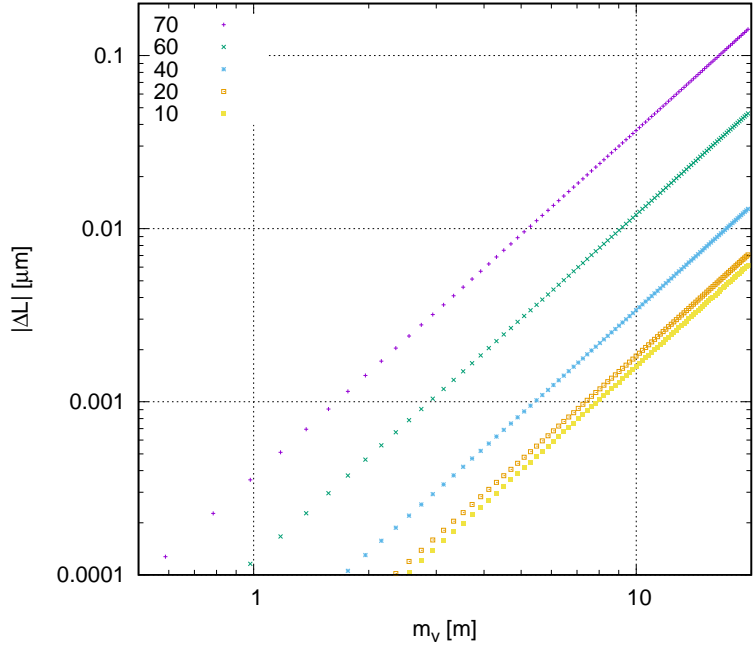}
\caption{Absolute value of the aberrations in the pupil plane with the Cassini
single-layer model, Eq.\ (\ref{eq.RCas}), with the parameters of Figure \ref{fig.Raberr}.}
\label{fig.ACas}
\end{figure}

The ray that will arrive at $(m_h,m_v)$ in the input pupil hits the top
of the atmosphere at a global coordinate
\begin{equation}
[\rho(0,0)+h]\left(\begin{array}{c}
\cos \alpha(m_h,m_v) \\
\sin\alpha(m_h,m_v) \sin\phi(m_h,m_v)\\
\sin\alpha(m_h,m_v) \cos\phi(m_h,m_v)
\end{array}\right)
\end{equation}
where $\phi$ of Eq.\ (\ref{eq.iplane}) fixes the incidence plane. The distance between this point
and $\vec m(m_h,m_v)$, multiplied by $n_0$, is the optical path length in air
(substituting $R\to R^{(C)}$ where necessary).
The angle of incidence of the ray traveling above the atmosphere in
the direction $(-1,0,0)$ is $\alpha$. 
Snell's law gives an exit angle $\alpha'$ of
$ \sin\alpha = n_0 \sin{\alpha'}$
relative to the local normal vector. 
\begin{equation}
\alpha-\alpha' = R^{(C)} = z-z_0.
\end{equation}
The incidence point has a geometric path difference $l$ to the telescope
\begin{equation}
[\rho(0,0)+h]
\left(\begin{array}{c}
\cos \alpha \\
\sin\alpha \sin\phi\\
\sin\alpha \cos\phi
\end{array}\right)
=
\rho(m_h,m_v)
\left(\begin{array}{c}
\cos z(m_h,m_v) \\
\sin z(m_h,m_v) \sin\phi(m_h,m_v)\\
\sin z\cos\phi(m_h,m_v)
\end{array}\right)
+l
\left(\begin{array}{c}
\cos R^{(C)}(m_h,m_v)\\
\sin R^{(C)}(m_h,m_v)\sin\phi(m_h,m_v)\\
\sin R^{(C)}(m_h,m_v)\cos\phi(m_h,m_v)
\end{array}\right).
\label{eq.Cassl}
\end{equation}
Adding the squares of the three Cartesian components gives
a quadratic equation for $l$ which is solved by
\begin{equation}
l
=
 -\rho(m_h,m_v) \cos z_0(m_h,m_v)+\sqrt{ [\rho(0,0)+h]^2-\rho^2(m_h,m_v) \sin^2 z_0(m_h,m_v)}
.
\end{equation}
The optical path length is this path through air plus a
stretch in vacuum which we let start at a reference plane tangential to the top of the atmosphere:
\begin{equation}
L^{(C)} = n_0 l +[\rho(0,0)+h](1-\cos \alpha).
\end{equation}
We are only interested in path differences, so we drop the constant $\rho(0,0)+h$, then
substitute the first component of the vector in Eq.\ (\ref{eq.Cassl}):
\begin{eqnarray}
L^{(C)}&=& [n_0-\cos R^{(C)}(m_h,m_v)]\left[\sqrt{ [\rho(0,0)+h]^2-\rho^2(m_h,m_v) \sin^2 z_0(m_h,m_v)}-\rho(m_h,m_v) \cos z_0(m_h,m_v)\right]
\nonumber \\
&&
-\rho(m_h,m_v)\cos z(m_h,m_v).
\end{eqnarray}
Insertion of (\ref{eq.cosz}) gives
\begin{eqnarray*}
\Delta L^{(C)}
= 
&&
[n_0-\cos R^{(C)}(m_h,m_v)]\left[\sqrt{ (\rho(0,0)+h)^2-\rho^2(m_h,m_v) \sin^2 z_0(m_h,m_v)}-\rho(m_h,m_v) \cos z_0(m_h,m_v)\right] 
\nonumber \\
&&
-[n_0-\cos R^{(C)}(0,0)]\left[\sqrt{ (\rho(0,0)+h)^2-\rho^2(0,0) \sin^2 z_0(0,0)}-\rho(0,0) \cos z_0(0,0)\right]  +m_v\sin R^{(C)}(0,0)
\end{eqnarray*}
\begin{eqnarray}
= 
&&
[n_0-\cos R^{(C)}(m_h,m_v)]\left[\sqrt{ (\rho(0,0)+h)^2-\rho^2(m_h,m_v) \sin^2 z_0(m_h,m_v)}-\sqrt{\rho^2(m_h,m_v) -\rho^2(m_h,m_v)\sin^2 z_0(m_h,m_v)}\right] 
\nonumber \\
&&
-( (m_h,m_v)\to (0,0))
+m_v\sin R^{(C)}(0,0)
.
\label{eq.RCas}
\end{eqnarray}
Numerical evaluation like in Figure \ref{fig.ACas} shows that the effect $|\Delta L^{(C)}|$
in the Cassini model is roughly a tenth of the values $|\Delta L|$ of the exponential model.

\section{C++ implementation}\label{app.prog}

\subsection{Differential zenith angle}
A C++ program to compute $R$ as a function of $\chi_0$ and $K/\rho$
up to the order $\tan^{11} z_0$ with up to mixed 5th order for $\tan z_0$ and 
up to mixed 4th order for $\tan^3z_0$, $\tan^5z_0$ and higher is in
the \texttt{anc} directory of the arXiv database.
The source  files are \texttt{r.cxx}, \texttt{RefreExpo.cxx} and \texttt{RefrExpo.h}. 
The program is compiled with
\begin{verbatim}
g++ -o r -O2 r.cxx RefrExpo.cxx
\end{verbatim}
or
\begin{verbatim}
make
\end{verbatim}
or
\begin{verbatim}
autoreconf -i -s
configure --prefix=$PWD
make
\end{verbatim}
and is run with up to 4 options:
\begin{verbatim}
r [-C] [-c chi0] [-K K] [-r rho]
\end{verbatim}
where \texttt{chi0} is the value of $\chi_0$ at the telescope site, \texttt{K}
is the scale height in meters, and \texttt{rho} is the sum of the Earth radius
and telescope sea level altitude in meters. The switch \texttt{-C}
produces results with Cassini's model, using \texttt{K} for the layer height
and \texttt{rho} for the earth radius. The program produces a table
of $z_0$ in degrees and values of $R$ in arcseconds.

\subsection{Input Pupil Aberration}
Another C++ program to compute $\Delta L$ as a function of $\chi_0$, $K$, $\rho$,
$m_h$ and $m_v$ is also attached.
Source  files are named \texttt{raberr.cxx}, \texttt{RefreExpo.cxx}, \texttt{RefrExpo.h},
\texttt{M1aberr.cxx} and \texttt{M1aberr.h}; 
it is compiled with
\begin{verbatim}
g++ -o raberr -O2 raberr.cxx RefrExpo.cxx M1aberr.cxx
\end{verbatim}
or
\begin{verbatim}
make
\end{verbatim}
or
\begin{verbatim}
autoreconf -i -s
configure --prefix=$PWD
make
\end{verbatim}
and is called as
\begin{verbatim}
raberr [-c chi0] [-K K] [-r rho] [-R radius] [-1 Nstps] -z zenith
\end{verbatim}
where \texttt{chi0}, \texttt{K} and \texttt{rho} have the same meaning as above,
where \texttt{radius} is the entrance pupil (primary) radius in meters,
and where \texttt{zenith} is the zenith angle in degrees.

If the option \texttt{-1} is not used,
it generates a table with three values per line suitable to create Figure \ref{fig.ELT}: $m_h$ and $m_v$ in meters along concentric
circles of the azimuth of the entrance pupil, and $\Delta L$ in micrometers.

If the option \texttt{-1} is used, it generates a table with three values per
line suitable to create Figure \ref{fig.Raberr}: $m_v$
in meters for positive $m_v$, $|\Delta L|$ in micrometers, and the air 
mass $1/\cos(z_0)$. \texttt{Nstps} is the 
number of points on the $m_v$ axis with a maximum set by \texttt{radius}.

The Maple program \texttt{CassAber.mp} computes a table of values
like shown in Figure \ref{fig.ACas}. 
Because this is a straight implementation without any further precautions
against cancellations of digits, the internal number of digits has been
set to near-quad precision in the program. 
The procedure \texttt{Cassgam}
contains an extended version of the $\gamma_i^{(C)}$-expansions following (\ref{eq.Cchi1}).

\bibliography{all}

\end{document}